\documentclass[epj
]{svjour}
\usepackage{graphicx}
\begin{document}
\title{A Power Law for the Duration of High-Flow States in Heterogeneous Traffic Flows}
\author{Dirk Helbing\inst{1} \and Benno Tilch\inst{2}
}                     
%
%
\institute{$^{1}$ETH Zurich, UNO D11, Universit\"atstr. 41, 8092 Zurich, Switzerland\\
$^2$II. Institut f\"ur Theoretische Physik, Pfaffenwaldring 57, 70550 Stuttgart, Germany}
\date{Received: date / Revised version: date}
%
\abstract{
We study the duration of ``high-flow states'' in freeway traffic, defined as the time periods for which traffic flows exceed a given flow threshold. Our empirical data are surprisingly well represented by a power law. Moreover, the power law exponent is approximately 2, in dependently of the chosen flow threshold. In order to explain this discovery, we investigate a simple theoretical model of heterogeneous traffic with overtaking maneuvers, which is able to reproduce both, the empirical power law and its exponent.
} 
\maketitle

\section{Introduction}

Efficient transport systems are needed to fulfil the requirements of industrialized societies. 
However, studies of traffic physicists have shown that the capacity of a freeway is reduced
by the breakdown of traffic flows \cite{Kerner94,Schadschneider,Review,Nagatani}, a phenomenon which is widely known as ``capacity drop''. Moreover, when the vehicle density increases, the traffic flow becomes metastable, i.e. a breakdown of traffic flow can be triggered by perturbations, if they exceed a critical perturbation threshold \cite{Kerner94,Persaud,Review}. At even higher densities, traffic flow becomes linearly unstable, and a breakdown is triggered by the slightest perturbation \cite{Sugiyama,Review}.
\par
Due to the dependence of traffic breakdowns on perturbations of the traffic flow, it is essential to know the characteristic properties of vehicle flows. While much attention has been paid to the measurement and explanation of the empirically observed wide scattering of congested traffic flows \cite{Kernersync}, the features of flows {\it before} the breakdown of free traffic have not found the attention they deserve. For example, the time period immediately before the breakdown is characterized by ``high-flow states'' (see Fig. \ref{hiflow}). These high-flow states are produced by small time gaps between subsequent vehicles, i.e. by vehicle platoons. Before we study these states, let us therefore shortly discuss some previous literature on vehicle platoons (see Ref. \cite{Review}).
\par\begin{figure}[htbp]
\caption[]{Empirical (a) Illustration of the capacity drop, and of widely scatterd flow-density data after the breakdown of traffic flow. (b) Illustration of the breakdown of the flow upstream of a bottleneck (50-vehicle averages) as a function of time. The flows before the breakdown are sometimes called ``high-flow states''. Congested traffic states can be characterized by falling below the flow $\rho V_{\rm sep}$ at a vehicle density $\rho$. The separating speed $V_{\rm sep}$ must be suitably chosen. Here, we set $V_{\rm sep} = 70$ km/h. (After \cite{Review}).}
\label{hiflow}
\end{figure}
The formation of platoons is typically a result of the fact that vehicles do not behave identically. Driver-vehicle behavior is rather heterogeneous, which is typically reflected by distributed model parameters.
The simplest models for heterogeneous transport are particle hopping models with {\em quenched
disorder}. For example, Evans \cite{Evans}, Krug and Ferrari \cite{Krug}, Karimipour \cite{Karimipour1,Karimipour2}, as well as Sepp\"al\"ainen and Krug \cite{Sepp}
study a simplified version of a model
by Benjamini, Ferrari, and Landim \cite{Ben}. It corresponds to the one-dimensional driven
lattice gas known as TASEP, but with particle-specific, constant jump rates $q_\alpha$.
When overtaking is not allowed, 
Krug and Ferrari \cite{Krug} find a sharp phase transition between a low-density regime, where
all particles are queueing behind the slowest particle, and a high-density regime, where
the particles are equally distributed. 
Note that the slow particles ``feel free traffic'' until the critical density is reached,
at which traffic flow becomes unstable
(cf. the truck curve in Fig.~\ref{carstrucks}). 
\par\begin{figure}[htbp]
\caption[]{Average speed of vehicles as a function of the vehicle density in (a) the right lane and (b) the left (fast) lane (after \cite{Moving}). The curves have been determined from single vehicle data of the Dutch freeway A9 close to Amsterdam. Note that the average velocity of trucks stays constant upto the critical density, where the speed of cars and trucks drops simulaneously and becomes more or less the same. This incidates that trucks ``feel'' free traffic conditions at all vehicle densities upto the occurence of the transition to congested traffic.}
\label{carstrucks}
\end{figure}
Close to the critical density, where the traffic flow becomes unstable, the 
growth of particle clusters (``platoons'') is characterized by a power-law coarsening.
If particles move ballistically with individual velocities $v_\alpha$ and
form a platoon when a faster particle reaches a slower one,
the platoon size $n_{\rm pl}(t)$ grows according to
\begin{equation}
 n_{\rm pl}(t) \sim t^{(\gamma+1)/(\gamma+2)} \, ,
 \label{obvious}
\end{equation} 
where the exponent $\gamma$ characterizes the distribution $P_0(v)
\sim (v - v_{\rm min})^{\gamma}$ of free velocities
in the neighborhood of the minimal desired velocity $v_{\rm min}$
\cite{BenNaim,Krapivsky1,Krapivsky2,Krapivsky3,Nagatani1,Nagatani2,Emmerich}.
Beyond it, the differences among fast and slow particles become irrelevant, 
because there is so little space that {\em all} particles have to move slower than preferred. 
\par
Platoon formation and power-law coarsening has also been found in microscopic models
with parallel update \cite{BenNaim,Nagatani2,Krapivsky1,Krapivsky2,Krapivsky3,Emmerich,Fukui,Nagatani3}. 
An example is the Nagel-Schreckenberg model with vehicle-specific slow-down probabilities 
\cite{Ktitarev,Knospe}.
\par
In real traffic, platoons remain limited in size. This is probably because of occasional possibilities for overtaking maneuvers on multi-lane roads. For a model for platoon size distributions see Islam and Consul \cite{Islam}.
\par
In the following, we will relate the distribution of platoon sizes with their growth dynamics,
and study the importance of overtaking maneuvers for both.
Before we start our theoretical considerations, however, Sec. \ref{Power} will present data of ``high-flow states'' and discusses their unexpected power-law statistics. Afterwards, Sec. \ref{NDeriv} will present a theoretical explanation, based on platoon formation due to overtaking maneuvers by slow vehicles. A discussion and outlook is given in Sec. \ref{Out}.

\section{Power Law Distribution of High-Flow States}\label{Power}

As indicated before, high traffic flows tend to be unstable, in particular if they exceed the outflow $Q_{
\rm out}$ from congested traffic not just temporarily due to a short fluctuation, but if they persist over a longer time period \cite{Persaud,Review}. It is, therefore, important to learn more about the statistics of ``high-flow states'', defined here by exceeding a given flow threshold $Q_{\rm thres}$ (which may be chosen different from $Q_{\rm out}$). 
\par
We were particularly interested in the {\it duration} of ``high flow states'' and analyzed single-vehicle data of the Dutch freeway A9 from Haarlem to Amsterdam in the Netherlands (see Ref. \cite{Springerbook} for details of the data). Specifically, we aggregated the data to obtain 2-minute averages of the flow $Q(x,t)$ as a function of time $t$ at a certain location $x$ of the freeway. Afterwards, we determined the time periods $\Delta t$ for which the flows stayed above a certain threshold $Q_{\rm thres}$. A representative example is shown in Fig. \ref{highflows}.
\par\begin{figure}[htbp]
\caption[]{Typical scaling law for the distribution of durations, for which the vehicle flow $Q(x,t)$ at the cross section $x$ of a freeway exceeds the threshold
$Q_{\rm thres} = 1400$ vehicles/h. The distribution follows a power law and has been determined from single vehicle data of the Dutch freeway A9 from Rottepolderplein to Badheuvedorp close to Amsterdam.}
\label{highflows}
\end{figure}
Similar pictures as for $Q_{\rm thres}=1400$ vehicles/h are found for other threshold values $Q_{\rm thres}$, but the data tend to be more noisy for large values of $Q_{\rm thres}$, as the typical durations of high-flow states become shorter. Generally, however, the probability distribution $P(\Delta t)$ of durations of high-flow states can be surprisingly well approximated by a power law
\begin{equation}
P(\Delta t) \sim (\Delta t)^{-\alpha}
\end{equation}
with an exponent $\alpha$. As the distributions $P(\Delta t)$ obviously depend on the chosen threshold value $Q_{\rm thres}$, it would be natural to assume that the respective power law exponents $\alpha$ depend on $Q_{\rm thres}$ as well. However, Fig. \ref{exponents} suggests that $\alpha \approx 2$, irrespective of the value of $Q_{\rm thres}$. Therefore, we have to find an explanation for both, the occurence of the power law for heterogeneous multi-lane traffic and the value $\alpha \approx 2$ of its exponent.
\begin{figure}[htbp]
\caption[]{Exponents of the power laws according to Fig. \ref{highflows}, for different thresholds $Q_{\rm thres}$. One can see that the exponents, fitted within the interval $\Delta t \in [6,50]$ min, are of the order of 2. The exponents have been determined from single vehicle data of the Dutch freeway A9 close to Amsterdam.}
\label{exponents}
\end{figure}

\section{Derivation of the Power Law Based on Overtaking Maneuvers}\label{NDeriv}

Our explanation starts with the hypothesis that high-flow states are produced by vehicle platoons, and that these are primarily a consequence of lasting overtaking maneuvers, particularly when one truck overtakes another one with a small relative velocity. Figure \ref{speedis} shows the speed distribution of cars and trucks in free traffic of low vehicle density. It can be seen that the speed of trucks varies around the applicable speed limit for trucks of 80 km/h, and that the distribution is quite narrow (i.e. speed differences are small). Furthermore, the speed distributions of cars and trucks can be reasonably well approximated by Gaussian distributions. A theoretical explanation of this is, for example, given in Ref. \cite{FPG}.
\par\begin{figure}[htbp]
\caption[]{Speed distribution of cars and trucks under free traffic conditions. Normal distributions fit the data reasonably well. The distributions have been determined from single vehicle data of the Dutch freeway A9 close to Amsterdam. Note that the left lane is the fast lane and that the speed limits for cars is 120 km/h on this freeway stretch, while it is 80 km/h for trucks.}
\label{speedis}
\end{figure}
It is natural that overtaking maneuvers of slow vehicles, e.g. overtaking trucks, constitute a moving
bottleneck for faster vehicles for some time. These faster vehicles will be queued up 
behind the slow ones (the trucks) and form a platoon. When a vehicle platoon passes a cross section of the road, the traffic flow is particularly high due to the small time gaps between its vehicles. In the following, we are interested in the theoretically expected statistics of time periods of high flows,
considering the Gaussian distribution of speeds.
\par
The growth of the queue length is determined by the arrival of faster vehicles at its end. Intuitively, 
the number of vehicles in the platoon grows proportionally with the
time required for the overtaking maneuver of the slow vehicles. Moreover, the 
period of high-flow states is expected to be proportional to the platoon length. In conclusion,
the time period $\Delta t$ of high flow states is determined by the time period required for
overtaking. 
\par
If $\Delta l_{\rm eff}$ denotes the effective distance over which an overtaking maneuver
takes place and $\Delta v$ is the relative speed between an overtaking vehicle and the slower one it overtakes, we find the proportionality relation
\begin{displaymath}
 \Delta t = \frac{\Delta l_{\rm eff}}{\Delta v} \, .
\end{displaymath}
As a consequence, the distribution of $\Delta t$ is given by the distribution of $1/\Delta v$.
\par
As we are particularly interested in extended high-flow states, i.e. long platoons, we need to determine the distribution of speed differences between {\it slow} vehicles, we can focus on the speed distribution of {\it trucks}. From statistics, it is well-known that the difference of two identically, independently, Gaussian distributed variables is Gaussian distributed as well. That is, the relative speeds $\Delta v$ of slow vehicles follow a Gaussian distribution
\begin{equation}
 N(\Delta v) \, d\Delta v = \frac{1}{\sqrt{2\pi\theta}} \mbox{e}^{-(\Delta v)^2/(2\theta)} \, .
\end{equation}
\par
From this fact, we can derive the distribution of the variable $y = 1/\Delta v$ by application of the
appropriate transformation. Considering 
\begin{equation}
 \frac{dy}{d\Delta v} = - \frac{1}{\Delta v^2} = -y^2 \, ,
\end{equation}
we find
\begin{equation}
 N(\Delta v) \, d\Delta v = - N(\Delta v) \frac{1}{y^2} dy \, , 
\end{equation}
where the minus sign is compensated for by integration from small to large values of $y = 1/\Delta v$ rather than vice versa. Therefore, we finally get the distribution
\begin{equation}
 P(y)\, dy = \frac{1}{\sqrt{2\pi\theta}} \, \frac{1}{y^2}\mbox{e}^{-1/[2\theta y^2]} \, dy \, .
\end{equation}
In the limit of small speed differences $\Delta v$, i.e. large values of $y$, we finally obtain the power law
\begin{equation}
 P(y) \sim y^{-2} \, .
\end{equation}
Furthermore, as $y= 1/\Delta v \sim \Delta t$, this implies the power-law distribution
\begin{equation}
 P(\Delta t) \sim (\Delta t)^{-2} \, ,  
\end{equation}
which is different from Eq. (\ref{obvious}).
Consequently, the durations $\Delta t$ of high-flow states should be distributed according to a
power law with exponent $-2$, which explains our empirical observations, see Fig. \ref{exponents}.

\section{Summary and Outlook}\label{Out}

In this paper, we have revealed a power law for the duration of high-flow states,
where ``high flow'' primarily means higher than some given threshold $Q_{\rm thres}$. 
Not only was it surprising to find that our empirical data could be approximated by a power law, but also that the power law exponent $\alpha$ was approximately 2, irrespective of the flow threshold $Q_{\rm thres}$. 
\par
Therefore, it was natural to look for an explanation of these surprising findings. Our hypothesis was that high-flow states occured due to vehicle platoons, and that these vehicle platoons would be caused by lasting overtaking maneuvers of slow vehicles, particularly of trucks. Based on the Gaussian distribution of the relative velocity of trucks, we could, in fact, derive the empirical power law and the exponent of 2.
\par
It should be added that vehicle platoons can cause the breakdown of traffic flows, as has been
studied in Ref. \cite{TranSci} in more detail. The typical scenario is a so-called ``boomerang effect'', where small perturbations in the flow move forward in the beginning, but turn their propagation direction after some time. According to Fig. \ref{overtakingtrucks}, the forwardly moving phase seems to be characterized by an increase in the {\it spatial extension} of the perturbation, i.e. growing vehicle platoons, while the upstream moving phase is characterized by a growing {\it amplitude} of the perturbation. This finally gives rise to a serious breakdown of traffic flow.
\begin{figure}
\caption[]{Illustration of a traffic breakdown on the German freeway A5 close to Frankfurt,
triggered by the so-called ``boomerang effect'',
which starts with a peak in the truck fraction (lower dotted
lines) (after \cite{TranSci}). For some subsequent time period, there is a peak in the flow
and a reduction of the speed, which indicates a dense vehicle cluster
queuing up behind overtaking trucks. While the cluster moves forward,
its spatial extension grows, indicating a growth of the platoon length. Finally, approximately one kilometer before the off-ramp at intersection Frankfurt North-West, which induces
many lane changing maneuvers because of the high off-ramp flow, the
velocity drops to values around 60~km/h, and a traffic jam travelling
against the driving direction appears. Afterwards, the reduction in
the average speed and flow increases with time, thereby causing a
growth in the amplitude of perturbation.}
\label{overtakingtrucks}
\end{figure}

\begin{acknowledgement}
The authors would like to thank Rudolf Sollacher for stimulating the empirical investigation of the characteristics of high-flow states and the Dutch Ministry of Transport, Public Works and Water Management, in particular Henk Taale, for providing the data of the Dutch freeway A9. The data of the German freeway A5 were provided by the Hessisches Landesamt f\"ur Stra{\ss}en- und Verkehrswesen.
Peter Felten and Mehdi Moussaid fitted the Gauss curves to the data of Fig. 5.
\end{acknowledgement}


\begin{thebibliography}{99}
\bibitem{Kerner94}
B. S. Kerner and P. Konh\"auser, 
Structure and parameters of clusters in traffic flow,
{\it Phys. Rev. E} {\bf 50}, 54--83 (1994).

\bibitem{Schadschneider} 
D. Chowdhury, L. Santen, and A. Schadschneider,
Statistical physics of vehicular traffic and some related systems.
{\it Physics Reports} {\bf 329}, 199 (2000).

\bibitem{Review} 
D. Helbing, Traffic and related self-driven many-particle systems.
{\it Reviews of Modern Physics} {\bf 73}, 1067--1141 (2001).

\bibitem{Nagatani}
T. Nagatani, The physics of traffic jams.
{\it Reports on Progress in Physics} {\bf 65}, 1331--1386 (2002).

\bibitem{Persaud}
B. Persaud, S. Yagar, and R. Brownlee, 
Exploration of the breakdown phenomenon in freeway traffic,
{\it Transpn. Res. Rec.} {\bf 1634}, 64--69 (1998).

\bibitem{Sugiyama}
Y. Sugiyama, M. Fukui, M. Kikuchi, K. Hasebe, A. Nakayama, K. Nishinari, S.-i. Tadaki, and S, Yukawa,
Traffic jams without bottlenecks---experimental evidence for the physical mechanism of the formation of a jam, {\it New Journal of Physics} {\bf 10}, 033001 (2008).

\bibitem{Kernersync}
B. S. Kerner and H. Rehborn, 
Experimental properties of complexity in traffic flow
{\it Phys. Rev. E} {\bf 53}, R4275--R4278 (1996).

\bibitem{Evans} 
M. R. Evans, 
Bose-Einstein condensation in disordered exclusion models and relation to traffic flow,
{\it Europhys. Lett.} {\bf 36}, 13--18 (1996).

\bibitem{Krug}
J. Krug and P. A. Ferrari, 
Phase transitions in driven diffusive systems with random rates,
{\it J. Phys. A: Math. Gen.} {\bf 29}, L465--L471 (1996).

\bibitem{Karimipour1}
V. Karimipour, 
A multi species asymmetric exclusion process, steady state 
and correlation functions on a periodic lattice,
{\it Europhys. Lett.} {\bf 47}, 304--310 (1999).

\bibitem{Karimipour2}
V. Karimipour, 
A remark on integrability of stochastic systems solvable 
by matrix product ansatz,
{\it Europhys. Lett.} {\bf 47}, 501--507 (1999).

\bibitem{Sepp}
T. Sepp\"al\"ainen and J. Krug, 
Hydrodynamics and platoon formation for a totally 
asymmetric exclusion model with particlewise disorder,
{\it J. Stat. Phys.} {\bf 95}, 525--567 (1999).

\bibitem{Ben}
I. Benjamini, P. A. Ferrari, and C. Landim, 
Asymmetric conservative processes with random rates
{\it Stoch. Process. Appl.} {\bf 61}, 181--204 (1996).

\bibitem{Moving}
D. Helbing and B. A. Huberman,
Coherent moving states in highway traffic. 
{\it Nature} {\bf 396}, 738--740 (1998).

\bibitem{BenNaim}
E. Ben-Naim, P.~L. Krapivsky, and S. Redner, 
{\it Kinetics of clustering in traffic flows}
{\it Phys. Rev. E} {\bf 50}, 822--829 (1994).

\bibitem{Krapivsky1}
E. Ben-Naim and P. L. Krapivsky, 
Stationary velocity distributions in traffic flows
{\it Phys. Rev. E} {\bf 56}, 6680--6686 (1997).

\bibitem{Krapivsky2}
E. Ben-Naim and P. L. Krapivsky,
Steady-state properties of traffic flows,
{\it J. Phys. A: Math. Gen.} {\bf 31}, 8073--8080 (1998).

\bibitem{Krapivsky3}
E. Ben-Naim and P. L. Krapivsky, 
Maxwell model of traffic flows,
{\it Phys. Rev. E} {\bf 59}, 88--97 (1999).

\bibitem{Nagatani1}
T. Nagatani, 
Self-organized criticality in {1D} traffic flow model with inflow or outflow,
{\it J. Phys. A: Math. Gen.} {\bf 28}, L119--L124 (1995).

\bibitem{Nagatani2}
T. Nagatani, 
Kinetics of clustering and acceleration in 1D traffic flow,
{\it J. Phys. Soc. Jpn.} {\bf 65}, 3386--3389 (1996).

\bibitem{Emmerich}
T. Nagatani, H. Emmerich, and K. Nakanishi, 
Burgers equation for kinetic clustering in traffic flow,
{\it Physica A} {\bf 255}, 158--162 (1998).

\bibitem{Fukui}
M. Fukui and Y. Ishibashi,
Traffic flow in 1D cellular automaton model including cars moving 
with high speed,
{\it J. Phys. Soc. Jpn.} {\bf 65}, 1868--1870 (1996).

\bibitem{Nagatani3}
T. Nagatani, 
Traffic behavior in a mixture of different vehicles,
{\it Physica A}, {\bf 284}, 405--420 (2000).

\bibitem{Ktitarev}
D. Ktitarev, D. Chowdhury, and D. E. Wolf,
Stochastic traffic model with random deceleration probabilities:
Queueing and power-law gap distribution,
{\it J. Phys. A: Math. Gen.} {\bf 30}, L221--L227 (1997).

\bibitem{Knospe}
W. Knospe, L. Santen, A. Schadschneider, and M. Schreckenberg, 
Disorder effects in cellular automata for two-lane traffic,
{\it Physica A} {\bf 265}, 614--633 (1999).

\bibitem{Islam}
M. N. Islam and P. C. Consul, 
The {C}onsul distribution as a bunching model in traffic flow,
{\it Transpn. Res. B} {\bf 25}, 365--372 (1991).

\bibitem{Springerbook}
D. Helbing,
{\it Verkehrsdynamik}
(Springer, Berlin, 1997).

\bibitem{FPG}
D. Helbing and M. Treiber,
Approximate Hamiltonian statistics in one-dimensional dissipative driven many-particle systems,
in preparation for {\it Eur. Phys. J. B} (2008).

\bibitem{TranSci}
M. Sch\"onhof and D. Helbing, Empirical features of congested traffic state and their implications for traffic modeling. Transportation Science {\bf 41}, 135--166 (2007).
\end{thebibliography}
\end{document}